\documentclass[prl,twocolumn,showpacs,floatfix]{revtex4}

\usepackage{bm}
\usepackage{graphicx}  
\usepackage{amsmath}     
\usepackage{amsfonts} 
\usepackage{amssymb}
\usepackage{amsthm}
\usepackage{dcolumn}

\def\Xint#1{\mathchoice
{\XXint\displaystyle\textstyle{#1}}%
{\XXint\textstyle\scriptstyle{#1}}%
{\XXint\scriptstyle\scriptscriptstyle{#1}}%
{\XXint\scriptscriptstyle\scriptscriptstyle{#1}}%
\!\int}
\def\XXint#1#2#3{{\setbox0=\hbox{$#1{#2#3}{\int}$}
\vcenter{\hbox{$#2#3$}}\kern-.5\wd0}}
\def\dashint{\Xint-}

\begin{document}
\title{Hamiltonian ratchet of conventional pure quasi-2D electron system}

\author{Eduard Takhtamirov}
\affiliation{Kotel{\textquoteright}nikov Institute of Radio Engineering and Electronics of RAS, Moscow, Russia}

\date{\today}

\begin{abstract}
We trace a simple mechanical model of a ratchet, and embed its setup in a conventional quasi-two-dimensional electron system in a semiconductor heterostructure. Expressed are two distinct microscopic mechanisms for such systems to serve as quantum ratchets producing the in-plane directed current from zero-mean time-dependent electric fields. The effective-mass swap ratchet is based on modulation of the mobility through alteration of the electron's effective mass. Modulation of the strength of the effective in-plane electric field marks the skin-effect ratchet. They can generate both linear and circular photocurrents, the later taking place in the dissipationless regime.
\end{abstract}

\pacs{05.60.-k, 
		72.40.+w,  
		73.40.-c, 
}
\maketitle

What we consider is not just yet another quantum ratchet: a quantum mechanical system capable of producing the directed particle or quasiparticle current when subjected to a periodic time-dependent force with zero mean value, see the reviews \cite{rev1,rev2,rev3}. The phenomenon named the photogalvanic or photovoltaic effect in solid state physics has long been known, ascribed to all-dimensional systems, housed for practical use, and swarming with various microscopic mechanisms, see the early review \cite{Belinicher} and references therein, and the recent \cite{Brouwer, Switkes, Ganichev, Tarasenko, Chepelianskii, Entin}. We venture to express apparently the simplest ratchet scheme allowing easy realization and plain optimization, yet described with very concise mathematics. A toy model, which we start with, immediately invokes as many as three different physical mechanisms accounting for appearance of the in-plane dc current when a quasi-2D electron system in a semiconductor heterostructure is subjected to ac electromagnetic field. One of them, the most involved, originating from the electron-impurity scattering has been suggested recently \cite{Tarasenko}. The remaining two mechanisms are very transparent and, what makes them actually noticeable,---they have a dissipationless regime. This brings such an electron system to a very rare class of the so-called Hamiltonian ratchets \cite{rev2,rev3}. Being not stuck to generally weak processes relying on inferior agents like impurities or phonons, these mechanisms can potentially result in very robust directed transport.

\begin{figure}[t]
\includegraphics [width=7.5cm]{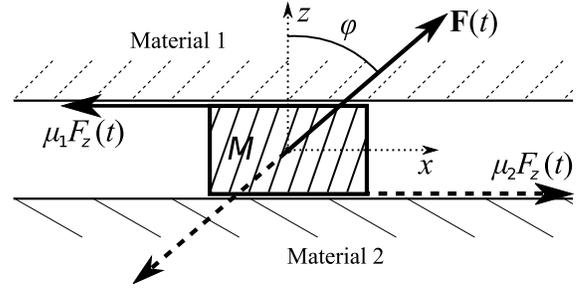}
\caption{
The quasi-2D friction ratchet: a block of mass $M$ put between flat walls of different materials and driven by an oblique zero-mean force $\mathbf F\left(t\right)$ attenuated by the friction force $\propto F_z\left(t\right)$. Solid arrowed lines: acting forces at ``positive'' drive. Dashed arrowed lines: acting forces at ``negative'' drive. Normal forces balancing $F_z\left(t\right)$ are not shown. If $\mu_2 > \mu_1$ and $\tan \phi > \mu_1$, the block's velocity $\dot x \left(t\right) \rightarrow +\infty$ as $t \rightarrow +\infty$.
} 
\label{fr.fig}
\end{figure}

Our core toy model is a mechanical arrangement of the friction ratchet type \cite{rev2}. We dub it the quasi-2D friction ratchet: a solid block of mass $M$ placed between flat walls of different materials and driven by a time-dependent force $\mathbf F\left(t\right)$ making an angle $\phi$ with a normal to the walls, see Fig. \ref{fr.fig}. Let constants $\mu_1$ and $\mu_2$ be the friction coefficients of the block's contacts with the two surfaces; for definiteness sake $\mu_2 > \mu_1$. The $x$-component of the force pulls the block while its $z$-component opposed with the normal force brings about friction. The Cartesian coordinates $x$, and $z$ refer to the rest reference frame, the $y$-coordinate be unaffected.
Consider a simplest zero-mean force, say, $\mathbf F\left(t\right) = \mathbf F_0 \sin \left( \Omega t\right)$ with the period $T= 2\pi \Omega^{-1}$; $F_{0x} = F_0 \sin \phi$ and $F_{0z} =F_0 \cos \phi$. The block is not frozen only if $\tan \phi > \min \left(\mu_1 , \mu_2 \right) = \mu_1$, upon which the equation of motion reads:
\begin{equation}
M \ddot {x}\left(t \right)  = F_x\left(t\right) - F_{\mathrm {att}}\left(t\right),
\label{block_eq}
\end{equation}
where $\ddot {x}\left(t \right) \equiv d^2 x\left(t \right)/dt^2$. The attenuation force $F_{\mathrm {att}}\left(t\right)$, expressed with the Heaviside $\Theta$-function, depends on $\phi$:
\begin{subequations}
\begin{equation}
F_{\mathrm {att}}\left(t\right) = \left[ \mu_1 \Theta \left( \sin \Omega t \right) + \mu_2 \Theta \left( -\sin \Omega t \right) \right] F_z\left(t\right)
\label{att1}
\end{equation}
if $\tan \phi > \mu_2$ with nonzero net forces for both half-cycles,
\begin{equation}
F_{\mathrm {att}}\left(t\right) =
\mu_1 F_z\left(t\right) \Theta \left( \sin \Omega t \right)+ F_x\left(t\right) \Theta \left(- \sin \Omega t \right)
\label{att2}
\end{equation}
\label{att}
\end{subequations}
if $\tan \phi < \mu_2$, with zero backlashes. Solving (\ref{block_eq}) with (\ref{att}), the increment of the block's momentum for the full period $0<t<T$ is $M\left(\dot x\left( T\right) - \dot x\left( 0\right) \right)= T \langle F\rangle$, where
\begin{equation}
\langle F\rangle = \frac 1 \pi \cdot
\begin{cases} 
 F_{0z} \left( \mu_2 - \mu_1 \right), & \tan \phi > \mu_2;\\
 F_{0x} - \mu_1 F_{0z}, & \tan \phi < \mu_2
\end{cases}
\label{deltap}
\end{equation}
can be viewed as the effective mean pulling force.

For the setup of our toy ratchet mimics a quasi-2D electron system lacking a center of symmetry, we may demand that such an electron system behave as a quantum ratchet. The affinity of the two arrangements is established with ease. Let the electrons with charge $-e<0$, effective mass $m^*$ and 2D-density $N_\mathrm s$ occupy a planar region restricted in the $z$-coordinate. If we apply the in-plane electric field $E_x\left(t\right)=\Re \left\{ E_{0x} \exp \left( i\Omega t\right)\right\} $, the Drude current density will be:
\begin{equation}
j_x\left(t\right)=\Re \left\{ \frac {e^2N_\mathrm s\tau} {m^*\left(1+i\Omega \tau \right)} E_{0x}{\mathrm e}^{i\Omega t}\right\},
\label{j_x}
\end{equation}
where $\tau$ is the momentum relaxation time. We are to find the conditions for the current (\ref{j_x}) to depend on the sign of the normally applied electric field $E_z$, which is to push the electron density $N\left( z\right)$ across the quantum well to the one wall or the other depending on the sign of $E_z$. We assume no vertical transport:
$\int_{-\infty}^{+\infty} N \left( z\right) dz = N_\mathrm s = \mathrm {const}$.
For that sign susceptibility, it must involve a quantity entering the right-hand side of Eq.~(\ref{j_x}). Then, as in our toy ratchet's scheme where we have arranged different conditions for the positive and negative drives to generate the nonzero mean motion, we will gain a directed current when we apply the modulating electric field $E_z\left(t\right)=\Re \{ E_{0z} \exp \left( i\Omega t\right)\}$ with the proper phase and the same frequency as the pulling in-plane field $E_x\left(t\right)$. An asymmetric in the $z$-coordinate distribution of scatterers, e.g., an asymmetric impurity doping profile makes $\tau$ depend on the sign of $E_z$. This is the first microscopic mechanism for the quantum implementation of our toy ratchet: the scattering quantum ratchet suggested in \cite{Tarasenko}. We will not discuss it here. The second mechanism, the effective-mass swap, relies on dependence of the effective mass $m^*$ on the sign of $E_z$. For this mechanism, having an asymmetric quantum well, besides, we must take into account a difference in the in-plane effective masses for different 2D subbands. In the simplest single-band case the difference appears, e.g., as we allow for position dependence of the effective mass and the non-parabolicity of the electron spectrum of the host material, their impacts being generally of the same order in magnitude \cite{Takhtamirov}. Finally, we closely inspect the last not unalterable quantity entering Eq.~(\ref{j_x})---the amplitude of the pulling field $E_{0x}$---for the third microscopic mechanism. It does not actually require an asymmetry of the quasi-2D electron system itself when the system is asymmetrically irradiated: it is the skin effect that brings in the asymmetric $z$-dependent field's amplitude $E_{0x}\left( z\right)$. This skin-effect ratchet operates by virtue of modulation of the effective pulling field rather than the electron's mobility: crudely, the electron density is forced to plunge into the region of stronger field, say, for the positive half-cycle and emerge for negative creating the nonzero average current.

These microscopic mechanisms turning a quasi-2D electron system into a quantum ratchet are independent and can be formalized separately. Let us start with the effective-mass swap ratchet and consider the single band one-electron Hamiltonian:
\begin{equation}
H_{10} = \frac {p_z^2} {2m^*_0} + U\left(z\right) + \hat {\mathcal M}\frac {p_x^2+p_y^2} 2 ,
\label{H10}
\end{equation} 
where $\left( p_x,p_y,p_z\right)=\mathbf p$ is the momentum operator, $m^*_0$ is a constant determining the effective mass for the quantized motion, $U\left(z\right)$ is an asymmetric function for the potential energy of the electron. The operator $ \hat {\mathcal M}$
provides a different in-plane effective mass for each of the 2D subbands. For a heterostructure of cubic-lattice semiconductors grown along the $\left[ 001\right]$ direction we have:
\begin{equation}
\hat {\mathcal M} = \hat {\mathcal M}\left( z, p_z\right) = \frac 1 {m^* \left( z\right)} + \left(4 \alpha + 2 \beta \right)p^2_z,
\label{effectivemass}
\end{equation}
where $m^* \left( z\right)$ is the position-dependent effective mass, with $\vert m^* \left( z\right) - m^*_0\vert \ll m^*_0$ being an applicability condition for the single band approximation, while $\alpha$ and $\beta$ are the weak non-parabolicity parameters of the host material, see, e.g.,~\cite{alfa-beta}.
The spin-degenerated eigenvalues $\varepsilon_{n}^k$ and normalized eigenfunctions $\vert n\mathbf k \rangle$ of (\ref{H10}) are listed with the subband index $n=1,2,\ldots$ and quasi-wavenumber $\mathbf k= \left(k_x,k_y\right)$. In what follows we will use the approximations:
\begin{equation}
\langle {\mathbf r}\vert n\mathbf k  \rangle = \frac {\mathrm e^{ik_xx+ik_yy}} {2\pi}\, \xi_n \left( z\right), \ \varepsilon_{n} ^k = \varepsilon_{n}^0 +  \hat {\mathcal M}_{nn}\frac {\hbar^2 k^2} 2 ,
\label{energy}
\end{equation}
where $\hat {\mathcal M}_{nn} \equiv \langle n 0 \vert \hat {\mathcal M} \vert n 0 \rangle$. In thermodynamic equilibrium the system is described with the one-particle density matrix $\rho_0$ subject to $\mathrm {Tr} \left( \rho_0 \right) = 1$:
\begin{equation}
\langle  n \mathbf k \vert \rho _0 \vert  n' \mathbf k' \rangle = 2N_\mathrm s^{-1} f_n^k \ \delta _{n n'}
\delta \left( \mathbf k - \mathbf k ' \right),
\label{ro0}
\end{equation}
$\delta_{a b}$ is the Kronecker delta, $\delta \left( x \right)$ is the Dirac delta function, and $f_n^k = f\left( \varepsilon_n^k \right)$ is the Fermi distribution function.

To try our system, we apply the pulling $x$- and modulating $z$- components of the electric field: $E_{x(z)} \left( t\right) = E_{0x(z)} \exp \left( i\Omega t\right) + \mathrm {c.c.}$, where $\mathrm {c.c.}$ stands for complex conjugate of the preceding term. We imply the amplitudes $E_{0x(z)}$ are not too strong and the frequency $\Omega$ is not too high to involve states from other bands. However, $\Omega$ cannot be arbitrarily low for we approximate a real system with $\Omega \tau \gg 1$. It is convenient to have the pulling field expressed with the vector potential rather than scalar: $A_x\left(t \right) = ic \Omega ^{-1} E_{0x} \exp \left( i\Omega t\right) + \mathrm {c.c.}$, where $c$ is the velocity of light in vacuum, and either way for the modulating field, e.g., via the scalar potential: $\phi \left(z, t \right) = -z E_z \left(t \right)$. Now the Hamiltonian is:
\begin{equation}
H_{1} = H_{10} + \hat {\mathcal M} \frac e c A_x\left(t \right) p_x - e \phi \left(z, t \right),
\label{H1}
\end{equation} 
where we neglect the term proportional to $A^2_x\left(t \right)$ as insignificant.
The current density:
\begin{equation}
j_x = -e N_\mathrm s \mathrm {Tr} \left( \hat v _x \rho \right),
\label{j}
\end{equation}
where the velocity operator $\hat v_x = \hat v_x\left(t \right)$:
\begin{equation}
\hat v_x = i \hbar ^{-1} \left( H_1 x - x H_1 \right) = \hat {\mathcal M} \left( p_x + \frac e c A_x\left(t \right) \right).
\label{v_x}
\end{equation}
The density matrix $\rho = \rho \left(t \right)$ is the solution of the quantum kinetic equation $i\hbar \partial \rho /\partial t = \left(H_1\rho - \rho H_1 \right)$ with $\mathrm {Tr} \left( \rho \right)= 1$ and the initial condition $\rho = \rho _0$ for $t = -\infty$. In the lowest order in the applied electric field we have $\rho = \rho _0 +\delta \rho$: 
\begin{equation}\label{ro}\begin{split}
\langle  n \mathbf k \vert &\delta \rho \left(t \right) \vert  n' \mathbf k' \rangle
= 2N_\mathrm s^{-1}
\left(f_{n'}^k - f_n^k\right)\delta \left( \mathbf k - \mathbf k ' \right)  \\
&\times \Bigg\{  
\frac {eE_{0z} z_{n n '} + i e \hbar k_x \Omega ^{-1} E_{0x} \hat {\mathcal M} _{nn'} }
{\varepsilon _{n'}^k - \varepsilon _n^k - \hbar \Omega +i0 } \ \mathrm e^{i\Omega t}\\
&\quad + \frac {eE^*_{0z} z_{n n'} - i e \hbar k_x \Omega ^{-1} E^*_{0x} \hat {\mathcal M} _{n n '} }
{\varepsilon _{n '}^k - \varepsilon _n^k + \hbar \Omega +i0 } \ \mathrm e^{-i\Omega t} \Bigg \},
\end{split}\end{equation}
where $E^*_{0x(z)}$ is complex conjugate of $E_{0x(z)}$. Combining Eq.~(\ref{j})-(\ref{ro}), we have for the time-average:
\begin{equation}\label{jav}\begin{split}
\langle& j_x \rangle = \frac {ie^3 } {2\pi^2\Omega} \sum_{n,n '= 1}^{\infty}
z_{n n '}\hat {\mathcal M} _{n ' n } \int d^2 k
\left( f_n^k  - f_{n'}^k \right)\\
&\times \left\{
\frac {E_{0x}E^*_{0z}} {\varepsilon _{n '}^k - \varepsilon _n^k + \hbar \Omega +i0 }
- \frac {E^*_{0x}E_{0z}} {\varepsilon _{n '}^k - \varepsilon _n^k - \hbar \Omega +i0 }
\right\}.
\end{split}\end{equation}
Let us choose real subband zone-center wavefunctions: $\Im \left( \xi_n \left( z\right) \right) = 0$. We have $\langle j_x \rangle=\langle j_x \rangle _\mathrm c + \langle j_x \rangle _\mathrm l$ with the density of the circular photocurrent:
\begin{equation}\label{jav_circ}\begin{split}
\langle j_x \rangle _\mathrm c &=
\frac {2e^3\left(E^*_{0x} E_{0z} - \mathrm {c.c.}\right)} {i\pi \Omega}
\sum_{n,n'=1}^{\infty} z_{n n+n'}\hat {\mathcal M} _{n+n' n}\\
&\times \dashint _0^{+\infty}
\frac {
\left[ f_n^k - f_{n+n'}^k \right]
\left( \varepsilon_{n+n'}^k - \varepsilon_n^k \right)
}
{\left( \varepsilon_{n+n'}^k - \varepsilon_n^k \right)^2 - \left( \hbar \Omega\right)^2}\, k dk,
\end{split}\end{equation}
where $\dashint$ stands for the integral's principal value, and the linear photocurrent: $\langle j_x \rangle _\mathrm l=\langle j_x \rangle _{\mathrm l +} + \langle j_x \rangle _{\mathrm l-}$, where
\begin{equation}\label{jav_lin}\begin{split}
\langle j_x &\rangle _{\mathrm l \pm} =
\frac {e^3\left(E_{0x} E^*_{0z} + \mathrm {c.c.}\right)} {\pm \Omega }
\sum_{n,n'=1}^{\infty} z_{n n+n'}\hat {\mathcal M} _{n+n' n}\\
&\times \int _0^{+\infty}
\left( f_n^k - f_{n+n'}^k \right)
\delta \left( \varepsilon_{n+n'}^k - \varepsilon_n^k \pm \hbar \Omega \right)kdk.
\end{split}\end{equation}

For systematics of the circular and linear photocurrents see, e.g., \cite{Belinicher}. The linear photocurrent occurs only in the presence of dissipation: it follows the resonant absorption of ac field marking (\ref{jav_lin}). The circular photocurrent may flow in a dissipationless regime, which we meet in Eq.~(\ref{jav_circ}) with virtual intersubband transitions rather than real forging the effect. It is sweeping in the sense that it can be generated by an obliquely incident elliptically polarized electromagnetic wave with any reasonable $\Omega$. As a function of $\Omega$, the finite width of the peaks in the linear photocurrent and of the dispersive structures in the circular one is provided with the $k$-dependence of the difference $\varepsilon_{n+n'}^k - \varepsilon_n^k$ in (\ref{jav_circ}) and (\ref{jav_lin}): it is of the order of $\hbar k^2_\mathrm F \vert \hat {\mathcal M} _{n n} - \hat {\mathcal M} _{n' n'} \vert $ for $ \hbar \vert \Omega \vert \approx \vert \varepsilon_n - \varepsilon_{n'} \vert$, where $k_\mathrm F = \sqrt{2\pi N_\mathrm s}$ is the Fermi wavenumber. The width of these peculiarities and hence their amplitude is also considerably affected by the amount of imperfections unavoidably present in a real system. Another agent is large-scale fluctuation of the quantum well thickness leading to the inhomogeneous broadening of the peculiarities. If the last two broadening mechanisms dominate, we may just as well use $\varepsilon_{n+n'}^k - \varepsilon_n^k \approx \varepsilon_{n+n'}^0 - \varepsilon_n^0$ in (\ref{jav_circ}) and (\ref{jav_lin}) having only a qualitative result for $\hbar \Omega$ close to an intersubband energy: a peak in $\langle j_x \rangle_ \mathrm l$ and a dispersive structure in $\langle j_x \rangle _\mathrm c$. 

A product of two small parameters keeps $\langle j_x \rangle$ slim. The first parameter is $\Delta_1 = eE_{0z} \bar z /\bar \varepsilon$, where $\bar z$ is the typical non-diagonal matrix element $z_{nn'}$, and $\bar \varepsilon$ is the typical intersubband energy: $\bar \varepsilon \sim \varepsilon_2^0 - \varepsilon_1^0$. The second small parameter is $\Delta_2= m_0^* \bar {\mathcal M}$, where $\bar {\mathcal M}$ is the typical non-diagonal matrix element $\hat {\mathcal M} _{nn'}$. We do not aim to optimize an electron system's design for the directed trasport efficiency here, but make straightforward estimates for a plain arrangement. For GaAs infinite quantum well of thickness, say, $L = 10$nm with $m^*_0 = 0.6\cdot 10^{-28}$g, we evaluate: $z_{12} \approx 2$nm, $ \varepsilon_2^0 - \varepsilon_1^0 \approx 0.15$eV. So, $\Delta_1 \sim 10^{-4}$ in the field $E_{0z}$ as high as $100$V/cm, for thicker quantum wells the figure being not so pathetic as $\Delta_1 \propto L^3$. To acquire larger $\Delta_2$ we need the opposite: highly asymmetric thin quantum wells with large intersubband energies to activate the non-parabolicity term and/or deep penetration of the wavefunctions into the barriers for the position-dependent effective mass term to work, see (\ref{effectivemass}). To get a rough estimate, we consider the symmetric quantum well from the previous estimate and add some asymmetry with a weak electric field $E_\mathrm a$. It creates the additional potential energy $W\left( z\right) = eE_\mathrm a z$ treated as a perturbation. The wavefunctions of the states lose their even-odd parities to foster $\langle 1 0 \vert p_z^2 \vert 2 0 \rangle = -2m^*_0eE_\mathrm a z_{12}$ with the retained $z_{12} \approx 2$nm. We use the material parameters $\alpha \approx \beta \approx - 0.2\hbar^{-4}$eVnm$^{4}$ for GaAs \cite{alfa-beta}, and employ $E_\mathrm a=0.01$V/nm making sizeable $eE_\mathrm a L \sim \varepsilon_2^0 - \varepsilon_1^0$. From (\ref{effectivemass}) and the above estimates we arrive at $\Delta_2 \sim 0.04$. Finishing with the effective-mass swap ratchet, for $N_\mathrm s = 3\cdot 10^{11} \mathrm {cm} ^{-2}$ and $\vert \hbar \Omega \pm \bar \varepsilon\vert \sim \bar \varepsilon \sim 0.1$eV, in the fields $\vert E_{0x}\vert = \vert E_{0z} \vert = 100$V/cm, the current density (\ref{jav_circ}) is $\langle j_x \rangle _\mathrm c \sim 0.2\mu$A/cm at the frequency $\Omega/(2\pi) = 1$THz.

In contrast to the effective-mass swap, the skin-effect mechanism is to be regarded as essentially many-particle for the $z$-dependence of the pulling field is caused by the induced electron current itself. However, the appearance of the directed current is a weak non-linear effect tractable to the perturbative analysis. Accordingly, the skin-effect ratchet can also be described with a one-electron Hamiltonian entered by the field $\mathbf E \left( z,t\right)$ that must be found separately as a solution of the Maxwell's equations supplemented with the linear material relations. This still bears interesting variants. For example, while the finite scattering may play a key role in building the effective electric field, no scattering as such is necessary for the ratchet to operate once the field has been set up. But this ``impure'' origin of the field will eventually be revealed in appearance of the resonant circular and sweeping linear photocurrents!

Let $E_x \left( z, t\right) = \left( \mathcal E + S\left( z\right) E_{0x} \right) \exp \left( i\Omega t\right) + \mathrm {c.c.}$ be the pulling field. It has a homogeneous part $\mathcal E$ making no interest to us, and a $z$-dependent contribution proportional to a dimensionless function $S\left( z\right)$. The latter is real if the considered electron system is pure what will be implied below. The operating skin-effect ratchet is mastered with the single-electron Hamiltonian:
\begin{equation}
H_2 = \frac {{\mathbf p}^2} {2m^*_0} + U\left(z\right) + \frac {ep_x} {m^*_0c} A_x\left(z,t \right) - e \phi \left(z, t \right),
\label{H2}
\end{equation}
where $A_x\left(z, t \right) = S\left( z\right)\left\{ ic \Omega ^{-1} E_{0x} \exp \left( i\Omega t\right) + \mathrm {c.c.}\right\}$ and the constant effective mass are the only items that distinguish it from the Hamiltonian (\ref{H1}) for the effective-mass swap ratchet. In addition, $U\left(z\right)$ may be even in $z$ with all asymmetry concentrating in $S\left( z\right)$. The substitute $S\left( z\right)/ m^*_0$ for $\hat {\mathcal M}$ in the expressions (\ref{v_x})-(\ref{jav_lin}) completes the quantum mechanics frame for the skin-effect mechanism. To have the full-fledged ratchet, we are yet to produce the potentials entering (\ref{H2}). Let us consider a quasi-2D electron gas embedded in a medium with the lattice dielectric permeability $\kappa_0$. Let the system be exposed to a plane electromagnetic wave with its magnetic field $\mathbf B$ directed along the $y$-axis. It is this p-polarized or TM wave that provides us with both pulling  $E_x$ and modulating $E_z$ fields. The s-polarized or TE wave would yield only a pulling electric field making us borrow the modulating component from the TM wave to secure the directed current. The incoming radiation is specified with $B_y\left( x,z,t \right) = B_0 \exp\left( i\Omega t - i q_x x - i q_z z \right)$, where $\kappa_0 \Omega^2 /c^2 = q_x^2 + q_z^2$. Actually, to find the necessary distribution of the fields inside the electron gas, we must solve the system of Maxwell's equations with linear but non-local material relations, such as the relation between the electric displacement and electric field \cite{Dahl}. We will make it in a devoted publication. Here we assert that the ``bare'' skin field, which is calculated using the net local dielectric permeability $\kappa \left( z \right) = \kappa_0 - 4 \pi e^2 N \left( z \right)/m^*_0 \Omega^2$, may serve as a seed. The non-local contributions only modify the electric fields found using the local $\kappa \left( z \right)$. The modification may be very profound though, especially for $\Omega$ close to frequencies of the intersubband plasmon modes when drastic variation of the electric field's strength across the quantum well is naturally expected.

If we approximate our electron gas with the constant density $N = N_\mathrm s /L$ inside the quantum well $0<z<L$, and use the Maxwell's equations and the local $\kappa \left( z \right)$, we will have for the uniform in the $x$-coordinate part of the fields acting on electrons ($0<z<L$):
\begin{equation}
E_x \left( z , t \right)= E_1 \mathrm e^{i\Omega t - gz} +E_2 \mathrm e^{i\Omega t + gz},
\label{Ex}
\end{equation}
\begin{equation}
E_z \left( z , t \right)= - iq_x g^{-1} \left( E_1 \mathrm e^{i\Omega t - gz} - E_2 \mathrm e^{i\Omega t + gz}\right),
\label{Ez}
\end{equation}
where we assume the system does attenuate the field so that $\Omega_\mathrm p > \Omega$, where $\Omega_\mathrm p = \sqrt{4 \pi e^2 N / m^*_0 \kappa_0 }$ is the effective 3D plasma frequency, and $g = \sqrt{4 \pi e^2 N / m^*_0c^2 - q_z^2}$ is real.
For the amplitudes $E_1$ and $E_2$ we have as $g L \ll 1$:
\begin{equation}
E_1 +E_2 = \frac {2i \kappa_\mathrm e q_z +2\kappa_0 g^2L}
{2i\kappa_0\kappa_\mathrm eq_z + \kappa_0^2 g^2 L - \kappa_\mathrm e^2 q_z^2 L}
\cdot \frac {cq_z}{\Omega} B_0,
\label{E1+E2}
\end{equation}
\begin{equation}
E_1 - E_2 = \frac {2i \kappa_\mathrm e q_z g L +2\kappa_0 g}
{2i\kappa_0\kappa_\mathrm e q_z + \kappa_0^2 g^2 L - \kappa_\mathrm e^2 q_z^2 L}
\cdot \frac {cq_z}{\Omega} B_0,
\label{E1-E2}
\end{equation}
where $\kappa_\mathrm e = \kappa_0 \left( 1 - \Omega^2 _\mathrm p / \Omega^2\right)$.
Finally, from (\ref{Ex}) and (\ref{Ez}) we have as $g L \ll 1$: $\mathcal E = E_1 +E_2$, $S\left( z\right) = - gz$, $E_{0x} = E_1 - E_2$, and $E_{0z} = -i q_x g^{-1}\left( E_1 - E_2 \right)$. The associated small parameter, which plays the role of the parameter $\Delta_ 2$ for the effective-mass swap mechanism, is $\Delta_ 3 = g \bar z$. For a GaAs quantum well with $N_\mathrm s = 3\cdot 10^{11} \mathrm {cm} ^{-2}$ and $L = 10$nm it is $\Delta _3 \sim 10^{-3}$. This is the effect of the mere bare skin field yet to be amplified as the non-local contributions to the material relations are taken into account.

In conclusion, we have found two microscopic mechanisms turning plain quasi-2D electron systems into true quantum ratchets. Being devoid of a need for a dissipation mechanism, moreover, the most effective in the limit $\tau = \infty$, they are very tractable to optimization concerning the band structure arrangement. We considered a simplest system characterized with the single-band approximation. More complex ones utilizing the same principles will behave similarly. While the effective-mass swap ratchet is still a relatively sophisticated object, the skin-effect ratchet may probably be molded from a piece of any conductive material. 

I am grateful to Prof. V.A. Volkov for acquaintance with the quantum ratchet concept and interest to the work, supported by INTAS (Grant No. 05-1000008-8044) and RFBR (Grant No. 08-02-00206).


\begin{thebibliography}{99}

\bibitem{rev1}
F.~J\"ulicher, A.~Ajdari, and J.~Prost, Rev. Mod. Phys. {\bf 69}, 1269 (1997). 

\bibitem{rev2}
P.~Reimann, Phys. Rep. {\bf 361}, 57 (2002). 

\bibitem{rev3}
P.~H\"anggi and F.~Marchesoni, Rev. Mod. Phys. {\bf 81}, 387 (2009). 

\bibitem{Belinicher} V.I.~Belinicher and B.I.~Sturman, Sov. Phys. Usp. {\bf 23} 199 (1980), [Usp. Fiz. Nauk. {\bf 130} 415 (1980)]. 

\bibitem{Brouwer} P.W.~Brouwer, Phys. Rev. B {\bf 58}, R10135 (1998). 

\bibitem{Switkes}
M.~Switkes, C.M.~Marcus, K.~Campman, and A.C.~Gossard, Science {\bf 283}, 1905 (1999). 

\bibitem{Ganichev}
S.D.~Ganichev and W.~Prettl,  J. Phys.: Condens. Matter {\bf 15}, R935 (2003). 

\bibitem{Tarasenko}
S.A.~Tarasenko, JETP Lett. {\bf 85}, 182 (2007) [Pis'ma v ZhETF {\bf 85}, 216 (2007)]. 

\bibitem{Chepelianskii}
A.D.~Chepelianskii, M.V.~Entin, L.I.~Magarill, and D.L.~Shepelyansky,
Phys. Rev. E {\bf 78}, 041127 (2008). 

\bibitem{Entin}
M.V.~Entin and L.I.~Magarill, Phys. Rev. B {\bf 79}, 075434 (2009). 

\bibitem{Takhtamirov}
E.E.~Takhtamirov and V.A.~Volkov, JETP {\bf 89}, 1000 (1999) [ZhETF {\bf 116}, 1843 (1999)]. 

\bibitem{alfa-beta}
U.~Ekenberg, Phys. Rev. B {\bf 40}, 7714 (1989). 

\bibitem{Dahl}
D.A.~Dahl and L.J.~Sham, Phys. Rev. B {\bf 16}, 651 (1977). 

\end{thebibliography}
\end{document}